\documentstyle[pra,aps,amsfonts,multicol]{revtex}
\newcommand{\ba}[2]{\left(\begin{array}{#1}#2\end{array}\right)}
\newcommand{\tr}[1]{{\rm Tr}\left(#1\right)}
\newtheorem{theorem}{Theorem}

\newtheorem{conjecture}{Conjecture}
\newcommand{\qed}{\hfill$\Box$}

\draft
\title{The Lorentz singular value decomposition and its applications to pure states of 3 qubits.}
\author{Frank Verstraete\cite{FV}, Jeroen Dehaene\cite{JD} and Bart De
Moor\cite{BDM}\\
\small{Katholieke Universiteit Leuven,
Department of Electrical Engineering, Research Group SISTA}\\
\small{Kardinaal Mercierlaan 94, B-3001 Leuven, Belgium}}

\begin{document}

\pagestyle{plain} \pagenumbering{arabic}

\maketitle
\begin{abstract}
All mixed states of two qubits can be brought into normal form by
the action of SLOCC operations of the kind $\rho'=(A\otimes
B)\rho(A\otimes B)^\dagger$. These normal forms can be obtained by
considering a Lorentz singular value decomposition on a real
parameterization of the density matrix. We show that the Lorentz
singular values are variationally defined and give rise to
entanglement monotones, with as a special case the concurrence.
Next a necessary and sufficient criterion is conjectured for a
mixed state to be convertible into another specific one with a
non-zero probability. Finally the formalism of the Lorentz
singular value decomposition is applied to tripartite pure states
of qubits. New proofs are given for the existence of the GHZ- and
W-class of states, and a rigorous proof for the optimal
distillation of a GHZ-state is derived.
\end{abstract}


\begin{multicols}{2}[]
\narrowtext Local probabilistically reversible operations cannot
affect the intrinsic nature of the entanglement present in a
system. It is therefore interesting to look for the most general
local operations that wash out all the local information such that
only the non-local character remains. For a pure entangled state
of two qubits for example, it is well-known that it can be locally
transformed into a Bell state, which is indeed the only pure state
for which the local density operators do not contain any
information. Recently, a similar result was derived in the case of
mixed states \cite{Verstraetelorsvd}. The key ingredient of the
analysis was the existence of a Lorentz singular value
decomposition. In this report some new interesting properties of
this Lorentz singular value decomposition are derived and it is
shown how it is related to the existence of entanglement
monotones. Furthermore it leads to a criterion for a mixed state
to be convertible into another specific one with a non-zero
probability. It also leads to a transparent derivation of all
different normal forms for pure states of three qubits: a pure
state of three qubits is indeed uniquely defined, up to local
operations, by the two qubit density operator obtained by tracing
out one particle. It will be shown how the so-called GHZ- and
W-type states \cite{DVC00} arise. We will also give a rigorous
proof of the optimal way of distilling a GHZ-state, confirming the
results of Acin et al.\cite{AJDV00}.

\section{The Lorentz Singular Value Decomposition}
Let us consider a mixed state of two qubits and investigate the
orbit generated by probabilistically reversible SLOCC operations
of the kind \begin{equation} \rho'=(A\otimes B)\rho(A\otimes
B)^\dagger\label{LOCC}\end{equation} where $A,B$ are complex {\rm
2x2} matrices of determinant 1 and $\rho'$ is unnormalized. It
will turn out very convenient to work in the real $R$-picture
defined as
\[\rho=\frac{1}{4}\sum_{ij=0}^3R_{ij}\sigma_i\otimes\sigma_j\]
with $\{\sigma_i\}$ the Pauli spin matrices. As shown in
\cite{Verstraetelorsvd}, the determinant 1 SLOCC operations
(\ref{LOCC}) in the $\rho$-picture become proper orthochronous
Lorentz transformations in the $R$-picture: \[R'=L_A R L_B^T\]
Indeed, it is a well known accident that $SL(2,C)\simeq SO(3,1)$.
Note that local operations are very transparent in the
$R$-picture: operations by Alice amount to operations on the row
space of $R$, i.e. left matrix multiplication, while operations by
Bob result in column operations.

Let us now state a more refined version of the central theorem of
\cite{Verstraetelorsvd}:\footnote{In comparison with the original
theorem we introduced a more refined classification in the case of
non-diagonalizable $R$.} \begin{theorem} The {\rm 4x4} matrix $R$
with elements $R_{ij}=\tr{\rho\sigma_i\otimes\sigma_j}$ can be
decomposed as \[R=L_1\Sigma L_2^T\] with $L_1,L_2$ finite proper
orthochronous Lorentz transformations, and $\Sigma$ either of
unique real diagonal form
\[\Sigma=\ba{cccc}{s_0&.&.&.\\.&s_1&.&.\\.&.&s_2&.\\.&.&.&s_3}\]
with $s_0\geq s_1\geq s_2\geq |s_3|$ and $s_3$ positive or
negative, or of the form
\[
\Sigma=\left(\begin{array}{cccc} a & . & . & b \\ . & d & . & .
\\ . & . & d & . \\ c & . & . &
b+c-a\end{array}\right)\] with unique $a,b,c,d$ obeying one of
the four following relations: $(b=c=a/2)$; $\left((d=0=c)\wedge
(b=a)\right)$; $\left((d=0=b) \wedge (c=a)\right)$;
$\left((d=0)\wedge (a=b=c)\right)$.\label{t2}\end{theorem}

It is interesting to note that the Lorentz singular values are
the only invariants of a state under the SLOCC operations
(\ref{LOCC}).

The diagonalizable case is generic, and a diagonal $R$ corresponds
to a Bell-diagonal state. The existence of the non-diagonal normal
forms is a consequence of the fact that the Lorentz group is not
compact: these non-diagonal normal forms can only be brought into
diagonal form by infinite Lorentz transformations. Nevertheless,
even in those cases the Lorentz singular values are well defined
and given by:
\[[s_0,s_1,s_2,s_3]=[\sqrt{(a-b)(a-c)},\sqrt{(a-b)(a-c)},d,-d]\]
The corresponding normal form of the non-diagonalizable case in
the $\rho$-picture is given by:
\[\rho=\frac{1}{2}\ba{cccc}{b+c&.&.&.\\.&a-b&d&.\\.&d&a-c&.\\.&.&.&.}\]
The four distinct non-diagonal normal forms correspond to the
following states:
\begin{itemize}
\item $(b=c=a/2)$: these are rank 3 states (rank 2 iff
$(d=b=c)$) with the strange property that their entanglement
cannot be increased by any global unitary
operation\cite{Verstraetegenbell}.
\item $\left((d=0=c)\wedge (b=a)\right)$; $\left((d=0=b)
\wedge (c=a)\right)$: $\rho$ is separable and a tensor product of
the projector ${\rm diag}[1;0]$ and the identity.
\item $\left((d=0)\wedge
(a=b=c)\right)$: $\rho$ is the separable pure state ${\rm
diag}[1;0;0;0]$.
\end{itemize}

Let us now consider the case of a generic pure state for which we
always have the relation $s_0=s_1=s_2=-s_3$. This implies that $R$
itself is a Lorentz transformation, up to a constant factor that
will turn out to be the concurrence; the singlet state for example
is represented in the $R$-picture by $R={\rm diag}[1;1;1;-1]$.
This clarifies why filtering operations by one party is enough to
distill a singlet out of an non-maximally entangled pure state:
Alice or Bob can apply the filter corresponding to the Lorentz
transformation given by the inverse of $R$.

The success of the ordinary singular value decomposition is to a
large extent the consequence of the nice variational properties of
the singular values: the sum of the $n$ largest singular values is
equal to the maximal inner product of the matrix with whatever n
orthonormal vectors. Interestingly, a similar property holds for
the Lorentz singular values:

\begin{theorem}
The Lorentz singular values $s_0\geq s_1\geq s_2\geq |s_3|$ of a
density operator $R$ are variationally defined as:
\begin{eqnarray*}&&s_0 = \min_{L_1,L_2}\tr{L_1RL_2^T\ba{cccc}{1 & . & .
& .
\\ . & . & . &
. \\ . & . & . & . \\ . & . & . & .}} \\
&&s_0-s_1  =  \min_{L_1,L_2}\tr{L_1RL_2^T\ba{cccc}{1 & . & . & .
\\ . & 1 & . & . \\ . & . & . & . \\ . & . & . & .}} \\
&&s_0-s_1-s_2  =  \min_{L_1,L_2}\tr{L_1RL_2^T\ba{cccc}{1 & . & .
& . \\ . & 1 & . & . \\ . & . & 1 & . \\ . & . & . & .}} \\
&&s_0-s_1-s_2+s_3  = \min_{L_1,L_2}\tr{L_1RL_2^T}\end{eqnarray*}
where $L_1,L_2$ are proper orthochronous Lorentz transformations.
\label{tv}\end{theorem}

\emph{Proof:} We will give a proof for the fourth identity and the
other proofs follow in a completely analogous way. An arbitrary
Lorentz transformation can be written as \[L=\ba{cc}{1 & . \\ . &
V}\ba{cccc}{\cosh(\alpha) &
\sinh(\alpha) & . & . \\ \sinh(\alpha) & \cosh(\alpha) & . & . \\
. & . & 1 & . \\ . & . & . & 1}\ba{cc}{1 & . \\ . & W},\] where
$V$ and $W$ are orthogonal 3x3 matrices of determinant 1. There is
no restriction in letting $R$ be in normal diagonal form, and
therefore we have to find the minimum
of \[\tr{\ba{cccc}{\cosh(\alpha) & \sinh(\alpha) & . & . \\
\sinh(\alpha) & \cosh(\alpha) & . & . \\ . & . & 1 & . \\ . & . &
. & 1}\ba{cc}{1 & . \\ . & W}\Sigma\ba{cc}{1 & . \\ . & V}}\] over
all $V,W,\alpha$. Using the variational properties of the ordinary
singular value decomposition and the fact that the Lorentz
singular values are ordered, it is immediately clear that an
optimal solution will consist in choosing $W=I_3$, $V={\rm
diag}[-1;-1;1]$  and  $\alpha=0$ as $\cosh(\alpha)>\sinh(\alpha)$
and $s_0\geq s_1$. This ends the proof.\qed

\section{Local invariants versus entanglement}
Let us now investigate how the local invariant Lorentz singular
values are related to the concept of entanglement. Inspired by
theorem \ref{tv}, we define the quantities
$M_1(\rho)=\max(0,-(s_0-s_1-s_2))$ and
$M_2(\rho)=\max(0,-(s_0-s_1-s_2+s_3))$. As they are solely a
function of the non-local invariants of the density operator, we
suspect them to be related to the amount of entanglement present
in the considered state:

\begin{theorem}
$M_1(\rho)=\max(0,-(s_0-s_1-s_2))$ and
$M_2(\rho)=\max(0,-(s_0-s_1-s_2+s_3))$ are entanglement monotones.
\end{theorem}

\emph{Proof:} A quantity $M(\rho)$ is an entanglement monotone
\cite{vidalmono} iff its expected value decreases under the action
of every local operation. Due to the variational characterization
of the quantities $(s_0-s_1-s_2)$ and $(s_0-s_1-s_2+s_3)$, it is
immediately clear that both $M_1$ and $M_2$ are decreasing under
the action of mixing. It is therefore sufficient to show that for
every $A\leq I_2$, $\bar{A}=\sqrt{I_2-A^\dagger A}$, it holds that
\begin{eqnarray*}&&M_i(\rho)\geq\nonumber\\&& \tr{(A\otimes
I)\rho(A\otimes I)^\dagger}M_i\left(\frac{(A\otimes
I)\rho(A\otimes I)^\dagger}{\tr{(A\otimes I)\rho(A\otimes
I)^\dagger}}\right)\nonumber\\
&&+\tr{(\bar{A}\otimes I)\rho(\bar{A}\otimes
I)^\dagger}M_i\left(\frac{(\bar{A}\otimes I)\rho(\bar{A}\otimes
I)^\dagger}{\tr{(\bar{A}\otimes I)\rho(\bar{A}\otimes
I)^\dagger}}\right)\nonumber\end{eqnarray*} It should be clear
from the previous discussion that the following identity holds:
\[M_i\left(\frac{(A\otimes I)\rho(A\otimes
I)^\dagger}{\tr{(A\otimes I)\rho(A\otimes
I)^\dagger}}\right)=\frac{\det(A)M_i(\rho)}{\tr{(A\otimes
I)\rho(A\otimes I)^\dagger}}\] Indeed, $A/\sqrt{\det(A)}$
corresponds to a Lorentz transformations which cannot change the
Lorentz singular values. We therefore only have to prove that
$1\geq \det(A)+\det(\bar{A})$. Given the singular values
$\sigma_1,\sigma_2$ of $A$, this inequality is $1\geq
\sigma_1\sigma_2+(1-\sigma_1)(1-\sigma_2)$ which is trivially
fulfilled. \qed

Both $M_1$ and $M_2$, linear functions of the Lorentz singular
values, are therefore entanglement monotones that are analytically
calculable for mixed states: we did not use the concept of convex
roof formalism. It turns out that $M_2$ is equivalent to the
concurrence of a state as introduced by
Wootters\cite{Wootters,Verstraetelorsvd}:
\[C(\rho)=\max\left(0,-\frac{1}{2}(s_0-|s_1|-|s_2|+s_3)\right)=\frac{1}{2}M_2(\rho)\]
There is indeed a strong relation between the Lorentz singular
values and the eigenvalues $\{\lambda_i\}$ of the operator
$\sqrt{(\sigma_y\otimes\sigma_y)\rho^T(\sigma_y\otimes\sigma_y)\rho}$
introduced by Wootters\cite{Wootters}: \[\ba{c}{s_0 \\
s_1 \\ s_2 \\ s_3}=\ba{cccc}{1 & 1 & 1 & 1 \\ 1 & 1 & -1 & -1 \\
1 & -1 & 1 & -1 \\ -1 & 1 & 1 & -1}\ba{c}{\lambda_1 \\
\lambda_2 \\ \lambda_3 \\ \lambda_4}.\] Together with the
negativity \cite{VidalWerner,Verstraetelorsvd}, the above
entanglement monotones are the only ones for which an analytical
expression exists for whatever mixed two-qubit state.

The existence of entanglement monotones is interesting as it gives
necessary conditions for one state to be convertible into another
one by LOCC operations with probability 1. It is still an open
problem to find the sufficient conditions for the convertibility
of one mixed state into another one, although this was solved for
pure states \cite{Nielsen,Jonathan,Vidaljon}. If we relax the
constraints that the conversion has to succeed with unit
probability, the above formalism can give us some answers in the
case of mixed states. We have indeed shown that a generic state
can always be brought into Bell-diagonal form by the SLOCC
operations (\ref{LOCC}). The problem of one state to be
convertible into another one with a non-zero probability is
therefore reduced to the question whether one Bell diagonal state
can be transformed into another one. A Bell-diagonal state is
uniquely defined under the SLOCC operations (\ref{LOCC}), and
therefore the only local tool remaining is mixing. Numerical and
theoretical investigations indicate that a given Bell diagonal
state can only be converted into another one iff this last one is
a mixture of the original Bell-diagonal state with a separable
state, although a general proof has not been found. We conjecture
however that this is always true:
\begin{conjecture}
A two-qubit state $\rho_1$ can probabilistically be converted into
the state $\rho_2$ iff the Bell-diagonal normal form of $\rho_2$
is a convex sum of a separable state and the Bell-diagonal normal
form of $\rho_1$.
\end{conjecture}

It is clear that a trivial procedure exists to implement this
conversion with unit efficiency: mix the state with one that can
be locally made. Let us for example investigate whether the
Bell-diagonal $\rho_1$ with ordered eigenvalues $\{\lambda_i\}$
can be transformed into the Bell-diagonal $\rho_2$ with ordered
eigenvalues $\{\mu_i\}$. We can restrict ourselves to mixing with
separable Bell diagonal states lying on the boundary of the
entangled and separable states, and these have their largest
eigenvalue equal to $1/2$. Under the assumption of our
conjecture, conversion is possible iff the following constrained
system of equations in $x,y,z,t,P$ has a solution:
\begin{eqnarray*}\ba{cc}{1&0\\0&P_3}\ba{c}{\mu_1 \\ \mu_2 \\ \mu_3 \\ \mu_4}&=&(1-x)\ba{c}{\lambda_1\\
\lambda_2\\ \lambda_3\\ \lambda_4}+x\ba{c}{1/2\\y\\z\\t}\nonumber\\
(0\leq x\leq 1)&&(y,z,t\geq 0)\hspace{.3cm}
(y+z+t=1/2)\nonumber\end{eqnarray*} where $P_3$ is a $3\times 3$
permutation matrix. This system can readily be solved. Not
surprisingly, there is a close relation between majorization and
the above set of equations. Note also that a pure entangled state
can be converted probabilistically into whatever mixed state.

\section{Pure states of three qubits}
Let us now apply the Lorentz singular value decomposition on the
problem of determining the different classes under SLOCC
operations of pure three qubit states. We will not consider the
states that have a tensor product structure, as these are not
truly tripartite. Therefore we know that taking the partial trace
over whatever party will result in a rank 2 density operator of
two qubits. Due to theorem \ref{t2}, we know that this density
operator can be brought into one of two normal forms by SLOCC
operations of two parties: a Bell diagonal $\rho_1=p|\psi^+\rangle
\langle\psi^+|+(1-p)|\psi^-\rangle \langle\psi^-|$ with
$|\psi^+\rangle=(|00\rangle +|11\rangle)/\sqrt(2) ;
|\psi^-\rangle=(|00\rangle -|11\rangle)/\sqrt(2)$;  or a quasi
distillable $\rho_2=|\phi^+\rangle \langle\phi^+|+|00\rangle
\langle 00|$ with $|\phi^+\rangle=(|01\rangle
+|10\rangle)/\sqrt(2)$, but this last case is clearly not generic.

Purification of this second normal form directly leads to the
normal form $|100\rangle +|010\rangle +|001\rangle $, called the
W-state \cite{DVC00}. As shown in \cite{DVC00}, this state has
maximal two-partite entanglement distributed over all parties.

Purification of the Bell diagonal state results in \[|\psi\rangle=
a(|000\rangle +|110\rangle )+b(|001\rangle -|111\rangle ).\] If
the third party now applies the local operation
$A=\ba{cc}{a&b\\b&-a}^{-1}$, the GHZ state $|000\rangle
+|111\rangle $ with maximal true tripartite entanglement is
obtained.

Therefore we have given an alternative proof of the following
theorem by D\"ur, Vidal and Cirac\cite{DVC00}:
\begin{theorem}
Every pure tripartite entangled state can be transformed to either
the GHZ, or the W-state, by SLOCC operations.
\end{theorem}

Note that the SLOCC operations bringing a generic pure state to
the GHZ form are not unique but consist of a four-parameter
family. This happens because a pure tripartite state has 14
degrees of freedom and the three Lorentz transformations have 18
independent parameters. Indeed, if $A\otimes B\otimes
C|\psi\rangle=|GHZ\rangle $, then also
\begin{equation}\ba{cc}{a&0\\0&1/a}A\otimes\ba{cc}{b&0\\0&1/b}B\otimes\ba{cc}{1/ab&0\\0&ab}C|\psi\rangle=|GHZ\rangle
\label{blabla}\end{equation} with $a$ and $b$ complex numbers. The
single-copy distillation of a GHZ state is therefore not unique.
The probability by which an SLOCC operation (\ref{LOCC}) produces
the desired result can therefore be optimized such as to yield the
optimal single-copy distillation protocol. This optimal procedure
was previously found by Acin et al. \cite{AJDV00}; a rigorous
proof of the same result is presented in the appendix
\label{appA}.

A similar non-uniqueness exists in the case of distilling a state
of the class of W-states to the W-state. Indeed, if $A\otimes
B\otimes C|\psi\rangle=|W\rangle$, then the most general symmetry
operations are given by
\begin{eqnarray*}
&&A'\otimes B'\otimes C'|\psi\rangle=|W\rangle\\
&&A'=\ba{cc}{x&y\\0&1/x}A\\
&&B'=\ba{cc}{x&z\\0&1/x}B\\
&&C'=\ba{cc}{x&-x(y+z)\\0&1/x}C\end{eqnarray*} with $x,y,z$
arbitrary complex numbers. As every matrix can be written as the
product of a unitary matrix and an upper triangular matrix (this
is the so-called QR-decomposition), there are enough degrees of
freedom to make whatever one out of $A'$,$B'$ or $C'$ equal to a
unitary matrix. Numerical investigations reveal that one of these
three possibilities is also the optimal choice in the sense that
it will yield a distillation protocol that produces the W-state
with the highest possible probability. Therefore the optimal
distillation protocol of a W-state consists of two parties
applying a local filtering operation, while one party performs a
local unitary operation.

Finally, a natural question arises as how the previous results
generalize to the case of mixed states. Due to the fact that the
rank of a density matrix corresponding to a mixed state is higher
then 1, it is immediately clear that no SLOCC operations can exist
that yield a rank 1 GHZ-state. In \cite{lsvd} it is shown that the
optimal SLOCC operations in the case of mixed states are those
that produce a unique state from a given state such that all its
local density matrices are equal to the identity. Note that the
GHZ state is the only pure state with this property in the 3-qubit
case.

A second question concerns the generalization of the class of pure
W-states to mixed states: is the W-class of mixed states of
measure zero? This question was solved in \cite{ABLS01} (see also
\cite{VDD01b} for a simple derivation and generalizations), where
it was shown that the W-class of mixed state is not of measure
zero.

\appendix
\section{Optimal distillation of the GHZ state}
The most general local procedure of distilling a GHZ-state out of
a single copy of a pure state consists of a multi-branch protocol
in which different branches consist of different SLOCC operations
connected through equation (\ref{blabla}). There is no restriction
in taking all $\{A_i\},\{B_i\},\{C_i\}$ to have determinant 1, and
the SLOCC operations corresponding to each branch are of the form
\begin{eqnarray*}&&q_i A_i\otimes B_i\otimes
C_i|\psi\rangle=q_i\tau^{1/4}|GHZ\rangle\\
&&A_i=D^a_iA_0\hspace{.3cm} D_i^a=\ba{cc}{a_i&0\\0&1/a_i}\\
&&B_i=D^b_iB_0\hspace{.3cm} D_i^b=\ba{cc}{b_i&0\\0&1/b_i}\\
&&C_i=D^c_iC_0\hspace{.3cm} D_i^c=\ba{cc}{1/a_ib_i &0\\0& a_ib_i }
.\end{eqnarray*} Here $\tau$ is the 3-tangle
\cite{Wootters3tangle,lsvd} of $\psi$ and $q_i$ is a real
proportionality factor such as to assure that all the branches
together are implementable as a part of a POVM. This leads to a
necessary (but generally not sufficient) condition:
\begin{equation}\sum_i q_i^2 A_i^\dagger A_i\otimes B_i^\dagger
B_i\otimes C_i^\dagger C_i\leq I_8\label{neccond}\end{equation}
Each branch yields the GHZ-state with probability
$q_i^2\sqrt{\tau}$, and therefore the total probability is given
by $\sqrt{\tau}\sum_i q_i^2$, which has to be maximized. Due to
the condition (\ref{neccond}), an upper bound on this probability
can readily be derived. It will turn out that this upper bound is
achievable by a 1-branch protocol. Defining $p_i=q_i^2/(\sum_i
q_i^2)$, it holds that the total probability is bounded by
\begin{equation}\max_{\{A_i\},\{B_i\},\{C_i\}}\frac{\sqrt{\tau}}{\lambda_{\max}(\sum_i
p_i A_i^\dagger A_i\otimes B_i^\dagger B_i\otimes C_i^\dagger
C_i)}\label{cond1}\end{equation} where $\lambda_{\max}(X)$ denotes
the largest eigenvalue of operator $X$. An upper bound is
therefore obtained by minimizing this largest eigenvalue.
Therefore the standard techniques for differentiating the
eigenvalues of a matrix have to be used \cite{Horn}: given a
Hermitian matrix $X$, its eigenvalue decomposition $X=UEU^\dagger$
and its variation $\dot{X}$, then the variation on its eigenvalues
is given by $\dot{E}={\rm diag}\{U^\dagger\dot{X}U\}$. Here we
take
\begin{eqnarray*}
X&=&Z_0^\dagger\underbrace{\sum_i p_i D_i}_{D}Z_0\\
Z_0&=&A_0 \otimes B_0 \otimes C_0\\
D_i&=&|D^a_i|^2\otimes |D^b_i|^2\otimes |D^c_i|^2
\end{eqnarray*}
Note that varying the free parameters $\{a_i,b_i,p_i\}$ only
affects $D$ and not $Z_0$. In the case of an extremal maximal
eigenvalue all variations
$\dot{\lambda}_{\max}=\tr{\dot{E}P_{11}}$ with $P_{11}={\rm
diag}[1;0;0;0;0;0;0;0]$ have to be equal to zero:
\[\tr{(\delta D)Z_0UP_{11}U^\dagger Z_0^\dagger}=0\]
The following identities are easily verified:
\begin{eqnarray*}\frac{\delta
D}{\delta a_i}&=&\frac{2}{a_i}{\rm diag}[0,1,0,1,-1,0,-1,0]D_i\\
\frac{\delta D}{\delta b_i}&=&\frac{2}{b_i}{\rm
diag}[0,1,-1,0,0,1,-1,0]D_i\\
\frac{\delta D}{\delta \sqrt{p_i}}&=&2\sqrt{p_i}D_i\end{eqnarray*}
Therefore only the (real and positive) diagonal elements of
$Z_0UP_{11}U^\dagger Z_0^\dagger$ are of importance and let us
write them in the vector $z_0$. Similarly, we write the diagonal
elements of $D_i$ in the vector
$d_i=[1;|a_ib_i|^2;1/|b_i|^2;|a_i|^2;1/|a_i|^2;|b_i|^2;1/|a_ib_i|^2;1]$,
and the extremal relations become:
\begin{eqnarray}
\forall i:&&0= d_i^T{\rm diag}[0,1,0,1,-1,0,-1,0]z_0\nonumber\\
&&0= d_i^T{\rm diag}[0,1,-1,0,0,1,-1,0]z_0\nonumber\\
&&\mu = d_i^T z_0 \label{xyz}\end{eqnarray} where $\mu$ is the
Lagrange multiplier corresponding to the condition
$\sum_i(\sqrt{p_i})^2=1$. This forms sets of each time 3 equations
for 2 unknowns $a_i,b_i$, which can be shown to have exactly one
solution. Indeed, the first and second equation lead to
\begin{eqnarray}|a_i|^4&=&\frac{z_0(5)+z_0(7)/|b_i|^2}{z_0(4)+z_0(2)|b_i|^2}\nonumber\\
|b_i|^4&=&\frac{z_0(3)+z_0(7)/|a_i|^2}{z_0(6)+z_0(2)|a_i|^2}.\label{curv}\end{eqnarray}
Let us analyze how these equations behave. When $b_i\rightarrow 0$
then the solution of the first equation goes like $|a_i|\sim
1/\sqrt{|b_i|}$ and when $a_i\rightarrow 0$ then $|b_i|\sim
1/|a_i|^2$. Exactly the  opposite happens in the case of the
second equation, and due to this different asymptotic behaviour it
is assured that both curves cross and therefore at least one
solution exists for all (real positive) values of $z_0$. Moreover
there is always at most one solution. To prove this, we first note
that $|a_i|$ and $|b_i|$ can be scaled such that both curves cross
at the value $(1,1)$, and we call these rescaled variables $(x,y)$
and $\bar{z}_0$. The hyperbola $xy=1$ crosses both rescaled curves
(\ref{curv}) at $(1,1)$. Moreover it is trivial to check that the
hyperbola does not cross any of the rescaled curves anymore in the
first quadrant (this amounts to solving a quadratic equation), and
due to the asymptotic behaviour one curve lies below and the other
one above the hyperbola (except in $(1,1)$). Therefore both
rescaled curves have exactly one crossing. Therefore for all (real
positive) values in $z_0$, there is always exactly one real
solution for $|a_i|,|b_i|$, and as $z_0$ is independent of the
index $i$, all $|a_i|$ are equal to each other and the same
applies to the $|b_i|$. Therefore at most the phase of the
constants $\{a_i,b_i\}$ varies in different branches, and as this
amounts to local unitary operations we conclude that all branches
are equivalent and can be implemented by a one-branch protocol.
This implies that the upper bound (\ref{cond1}) can be reached.

In the case of a one branch protocol, the eigenvectors of $X$ can
be calculated analytically as $X$ becomes a tensor product of
$2\times 2$ matrices. Given particular determinant 1
transformations $A,B,C$ and taking $a,b$ to be real,  the
eigenvector $v$ corresponding to the largest eigenvalue of the
matrix $YY^\dagger$ with
$Y=\ba{cc}{a&0\\0&1/a}A\otimes\ba{cc}{b&0\\0&1/b}B\otimes\ba{cc}{1/ab&0\\0&ab}C$
happens to be $v=v_1\otimes v_2\otimes v_3$ with
\begin{eqnarray*}v_i&=&\ba{c}{\alpha_i\\-\beta_i+\sqrt{\alpha_i^2+\beta_i^2}}\\
\alpha_1&=&2\sqrt{A_{11}A_{22}-1}\hspace{1cm}\beta_1=A_{11}a^2-A_{22}/a^2\\
\alpha_2&=&2\sqrt{B_{11}B_{22}-1}\hspace{1cm}\beta_2=B_{11}b^2-B_{22}/b^2\\
\alpha_3&=&2\sqrt{C_{11}C_{22}-1}\hspace{1cm}\beta_3=C_{11}/(ab)^2-C_{22}(ab)^2\end{eqnarray*}
The conditions (\ref{xyz}) then imply that $\beta_i/\alpha_i$ is a
constant for all $i=1..3$:
\[\frac{A_{11}a^2-A_{22}/a^2}{\sqrt{A_{11}A_{22}-1}}=\frac{B_{11}b^2-B_{22}/b^2}{\sqrt{B_{11}B_{22}-1}}
=\frac{C_{11}/(ab)^2-C_{22}(ab)^2}{\sqrt{C_{11}C_{22}-1}}\] These
two equations have to be solved in the unknowns $a$ and $b$. $b$
can readily be written in function of $a$ through one of those,
and then a sixth order equation in the remaining unknown $a^2$
results. As shown above, only one solution corresponding to a
physical solution for $a$ and $b$ exists, and this solution can
easily  be solved numerically. The optimal local filtering
operations and the maximal probability of making a GHZ-state (an
entanglement monotone\cite{vidalmono}) can then easily be
calculated.\\
The solution obtained is completely equivalent to the one of Acin
et al. \cite{AJDV00}, although their proof did not include the
uniqueness of the solution and needed exhaustive numerical
calculations.\\
Note that the procedure outlined here is equally applicable to the
problem of distilling a GHZ-state in a higher dimensional system.

\acknowledgements FV acknowledges interesting discussions with K.
Audenaert, T. Brun, P. Hayden, J. Kempe, R. Gingrich, A. Acin, E.
Jane and G. Vidal. FV is very grateful to Hideo Mabuchi and the
Institute of Quantum Information at Caltech, where part of this
work has been done.

\end{multicols}

\begin{thebibliography}{99}
\bibitem[\star]{FV} frank.verstraete@esat.kuleuven.ac.be
\bibitem[\#]{JD} jeroen.dehaene@esat.kuleuven.ac.be
\bibitem[\$]{BDM} bart.demoor@esat.kuleuven.ac.be
\bibitem{Verstraetelorsvd} F. Verstraete, J. Dehaene and Bart De
Moor, Phys. Rev. A {\bf 64}, 010101(R) (2001).
\bibitem{Verstraetegenbell} F. Verstraete, K. Audenaert and B. De Moor,
Phys. Rev. A {\bf 64}, 012316 (2001).
\bibitem{Wootters} W. Wootters, Phys. Rev. Lett. {\bf 80}, 2245 (1998).
\bibitem{vidalmono} G. Vidal, J.Mod.Opt. {\bf 47}, 355 (2000).
\bibitem{VidalWerner} G. Vidal and R.F. Werner, quant-ph/0102117.
\bibitem{Nielsen} M. Nielsen, Phys. Rev. Lett. {\bf 83}, 436 (1999).
\bibitem{Jonathan} D. Jonathan and M. Plenio, Phys.Rev.Lett. 83, 3566
(1999).
\bibitem{Vidaljon} G. Vidal, D. Jonathan and M. Nielsen, Phys. Rev. A 62, 012304
(2000).
\bibitem{DVC00}W. D\"ur, G. Vidal and  J. I. Cirac, Phys. Rev.
A 62, 062314 (2000).
\bibitem{AJDV00} A.Acin, E. Jane, W. D\"ur and G. Vidal, Phys. Rev.
Lett. 85, 4811 (2000).
\bibitem{Wootters3tangle} V. Coffman, J. Kundu and  W.K. Wootters,
Phys. Rev. A {\bf 61}, 052306 (2000).
\bibitem{lsvd} F. Verstraete, J. Dehaene and B. De Moor,  quant-ph/0105090.
\bibitem{ABLS01} A. Acin, D. Bru\ss, M. Lewenstein and A.
Sanpera, Phys. Rev. Lett. 87, 040401 (2001).
\bibitem{VDD01b} F. Verstraete, J. Dehaene and B. De Moor, quant-ph/0107155.
\bibitem{Horn} R. Horn and C. Johnson, {\em Matrix Analysis}, Cambridge University Press
(1985); R. Horn and C. Johnson, {\em Topics in Matrix Analysis},
Cambridge University Press (1991).


\end{thebibliography}
\end{document}